 \documentclass[12pt,preprint]{aastex}

\usepackage{color}
\usepackage{ulem}
\usepackage{amsmath}

\shorttitle{magnetic field in luminous AGNs  }
\shortauthors{Liu et al.}

\begin{document}


\title{Revisiting the structure and spectrum of the magnetic-reconnection-heated corona in luminous AGNs}


\author{J.Y. Liu\altaffilmark{1,2}, E. L. Qiao\altaffilmark{3} and B. F. Liu\altaffilmark{3}}
\altaffiltext{1}{National Astronomical Observatories/Yunnan Observatory, Chinese Academy of Sciences,
Kunming 650011, China }\email{ljy0807@ynao.ac.cn, qiaoel@nao.cas.cn, bfliu@nao.cas.cn}
\altaffiltext{2}{ Key Laboratory for the Structure and Evolution of Celestial Objects, Chinese Academy of Sciences, Kunming 650011, China}

 \altaffiltext{3}{National Astronomical Observatories, Chinese Academy of
Sciences, Beijing 100012, China}

\begin{abstract}
It is believed that the hard X-ray emission in the luminous active galactic nuclei (AGNs) is from the hot corona above the cool accretion disk. However, the formation of the corona is still debated. Liu et al. investigated the spectrum of the corona heated by the reconnection of the  magnetic field generated by dynamo action in the thin disk and emerging into the corona as a result of buoyancy instability. In the present paper, we improve this model to interpret the observed relation of the hard X-ray spectrum becoming softer at higher accretion rate in luminous AGNs. The magnetic field is characterized by $\beta_{\rm 0}$, i.e., the ratio of the sum of gas pressure and radiation pressure to magnetic pressure in the disk ($\beta_{\rm 0}=(P_{\rm  g,d}+P_{\rm r,d})/P_{\rm B}$). Besides, both the intrinsic disk  photons and reprocessed photons by the disk are included as the seed photons for inverse Compton scattering. These improvements are crucial for investigating the effect of magnetic field on the accretion disk-corona when it is not clear whether the radiation pressure or gas pressure dominates in thin disk. We change the value of $\beta_{\rm 0}$ in order to constrain the magnetic field in the accretion disk. We find that the energy fraction released in the corona ($f$) gradually increases with the decrease of $\beta_{\rm 0}$ for the same accretion rate. When $\beta_{\rm 0}$ decreases to less than 50, the structure and spectrum of the disk-corona is independent on accretion rate, which is similar to the hard spectrum found in Liu et al.(2003). Comparing with the observational results of the hard X-ray bolometric correction factor in a sample of luminous AGNs,
we suggest that the value of $\beta_{\rm 0}$ is about 100-200 for $\alpha=0.3$ and the energy fraction $f$ should be larger than $30\%$ for hard X-ray emission.
 \end{abstract}



\section{INTRODUCTION}

An active galactic nucleus (AGN) is a very compact region located at
the center of a galaxy, which can emit from radio to X-rays.  The
radiation from AGNs is believed to be powered by accreting the surrounding
matter onto the supermassive black hole. Observations indicate that
different types of AGNs may have different accretion modes. For the
low-luminosity AGNs (LLAGNs), roughly $L_{\rm bol} < 10^{44} \rm{~erg~
s^{-1}}$, due to the low radiative efficiency, the radiation is
generally believed to be dominated by a faint, radiatively
inefficient accretion flow (RIAF; e.g., Narayan \& Yi 1994, 1995a, 1995b;
Quataert et al. 1999; Ho 2008; Yuan \& Narayan 2014). The case of luminous AGNs, mainly including quasars and bright Seyfert galaxies, 
is different. The spectral energy distribution (SED) of luminous
AGNs can be characterized by different components: a `big blue bump'
in optical-to-ultraviolet (UV) band, which is often explained by a
geometrically thin, optically thick, accretion disk extending down
to the innermost stable circular orbits (ISCO; Shakura \& Sunyaev
1973; Shields 1978; Malkan \& Sargent 1982; Elvis et al. 1994;
Kishimoto et al. 2005; Shang et al. 2005); a soft X-ray
excess, whose origin is still unclear (Done et al. 2007); and a power-law hard X-ray emission, which is believed to be
produced by the inverse Compton scattering of the soft photons from
the accretion disk in a hot corona above (e.g., Svensson \& Zdziarski 1994; Magdziarz et al. 1998; Chiang 2002; Vasudevan \& Fabian 2009). In luminous AGNs, the hard X-ray spectrum is often
described by  a power law with $\Gamma_{\rm2-10~keV}\sim1.9$. Meanwhile, it is found that
there is a positive correlation between $\Gamma_{\rm2-10~keV}$ and Eddington ratio $\lambda_{\rm Edd}$ ($\lambda_{\rm Edd}= L_{\rm bol}/L_{\rm Edd}$ and $L_{\rm Edd}=1.26\times10^{38}~(M_{\rm BH}/M_{\rm \odot})~\rm{ erg~s^{-1}}$), 
Whereas a negative correlation is found between $\Gamma_{\rm2-10~keV}$ and $\lambda_{\rm Edd}$ in
LLAGNs
(Gu \& Cao 2009; Yang et al. 2015). In luminous AGNs, the hard X-ray luminosity is also used to
estimate the bolometric luminosity with the hard X-ray bolometric correction factor $L_{\rm bol}/L_{\rm 2-10~keV} \approx 20-150$
for $\lambda_{\rm Edd}>0.1$ (Wang et al. 2004; Vasudevan \& Fabian 2007, 2009; Zhou \& Zhao 2010; Fanali et al. 2013).

The formation of coronas in AGNs is still unclear. Previous works have revealed that a corona can be fed by the evaporation of matter
from an underlying cool disk (e.g., Meyer \& Meyer-Hofmeister 1994; Meyer
et al. 2000; Liu et al. 2002a; R\`{o}\.{z}a\`{n}ska \& Czerny 2000a, 2000b; Qian et al. 2007; Qiao \& Liu 2009).
In the framework of the disk evaporation model we mentioned above, if the mass accretion rate transferred from
the outermost region of the disk is less than
a predictive critical mass accretion rate $\dot M_{\rm crit} \sim 0.02 \dot M_{\rm Edd}$, the disk will be truncated
at a radius from the black hole. However, if the mass accretion rate transferred from the outmost region of the
disk is greater than $\dot M_{\rm crit}$,
the disk cannot be completely evaporated into the corona and extends down to
the ISCO of the black hole. In this model, generally, the corona is presumed to be heated by the viscous heating of the corona itself.
Meyer-Hofmeister et al. (2012) studied the strength of the corona for strong mass flow in the disk and found that the strong Compton cooling of the corona by the soft photons  from the disk makes the corona quickly
condense onto the disk. For a typical mass accretion rate of $0.1\dot M_{\rm Edd}$ transferred from the mostouter region of the disk,
the accretion rate in the corona fed by disk evaporation is very low, i.e., less than $0.001\dot M_{\rm Edd}$,
which is inconsistent with the observed strong X-ray emission in most of the luminous AGNs.

In order to resolve the energy-deficiency
problem, Liu et al. (2012) set the ratio ($f$) of the corona heating to the total gravitational
energy to be a free parameter and found that a relatively large value of $f$ is needed for luminous AGNs.
The more energy is released in the corona, i.e., the larger $f$, the harder the X-ray emits from the corona.
 For the corona heating, one of the possible mechanisms might be
magnetic reconnection. The magnetic field is generated by dynamo action in the accretion disk. Due to the buoyancy instability,
the magnetic flux loop emerges from the disk and reconnects
with other loops in the corona, thereby releasing the magnetic energy to heat the coronal plasma. The
energy is then radiated away through inverse Compton scattering (see as Tout \& Pringle 1992;
Di Matteo 1998; Miller \& Stone 2000; Merloni \& Fabian 2001; Liu et al. 2002b, 2003; Wang et al. 2004; R\`o\.za\`nska \& Czerny 2005; Cao 2009; Liu et al. 2012; You  et al. 2012;
Huang et al. 2014; Qiao \& Liu 2015).
Liu et al. (2002a, 2003) constructed a disk-corona model with the corona heated by the magnetic reconnection
and calculated the corresponding emergent spectra. The magnetic field strength is characterized by a parameter
$\beta \sim 1$ (with $\beta= P_{\rm g,d}/P_{\rm B}$, i.e., the ratio of gas pressure to magnetic pressure at the midplane of the disk). In their work, the disk is divided into two types, i.e., the
gas-pressure-dominated case and radiation-pressure-dominated case. They found two types of solutions corresponding to hard spectrum and soft spectrum. In the hard-spectrum solution, the energy fraction $f$ is nearly $1$,
and the hard X-ray spectrum index with $\alpha\sim1.1$ ($f_{\nu}\varpropto\nu^{-\alpha}$)
does not change with mass accretion rate. While, in the soft-spectrum solution, the energy fraction $f$ is nearly $0$,
implying that the X-ray emission is also nearly $0$. These two solutions only correspond to the two end points in the observed relation of the hard X-ray spectrum becoming softer at higher accretion rate in luminous AGNs. Even though there are composed solutions for moderate-luminosity, e.g., $\dot{M}=0.5\dot{M}_{\rm Edd}$ shown in the right panel of Fig.1 in Liu et al.(2003), the thin disk abruptly changes from being radiation pressure dominated to being gas pressure dominated at a radius, and the energy fraction also directly increases to 1 from 0. The distribution of the parameters in the corona, such as electron temperature, density, and the optical depth, are not continuous. However, the observed hard X-ray spectrum is softer at higher accretion rate, which means that the energy fraction smoothly decreases with the accretion rate. Obviously, the previous model cannot smoothly reproduce the observed fraction of the X-ray luminosity to the bolometric luminosity.

In the present work, in order to obtain a more self-consistent solution, we improve the model in Liu et al.(2002b, 2003) and investigate the effect of magnetic field on the structure and spectrum of such a disk-corona model. The disk is not divided into two types, and the magnetic pressure is assumed to be proportional to the sum of gas pressure and radiation pressure as characterized by magnetic parameter $\beta_{\rm 0}$ ($\beta_{\rm 0}=\frac{(P_{\rm g,d}+P_{\rm r,d})}{P_{\rm B}}$
). Besides, both the intrinsic disk radiation
and the backward corona Compton emission are always included for the
corona Compton cooling. These improvements will help us to investigate the properties of the accretion flows when it is not clear whether the radiation pressure or gas pressure dominates in the disk. These are crucial for smoothly changing of the energy fraction from 0 to 1 and predicting the observed hard X-ray emission in luminous AGNs. In order to constrain the magnetic field for certain viscous coefficient $\alpha$, we calculate the emergent spectrum of the model for different $\beta_{\rm 0}$ and
 compare the derived relation between hard X-ray bolometric correction and accretion rate with the observed relation in a sample of luminous AGNs. We suggest that the spectrum of the model with $\beta_{\rm 0}\sim\,100-200$ for viscous coefficient $\alpha=0.3$ is consistent with the observed result.

The structure of the paper is as follows: The model is presented in Section 2.
The numerical results for the structure and the emergent spectrum of the model are in Section 3.
The discussion and the conclusion are in Section 4 and Section 5 respectively.

\section{THE MODEL}
We adopt a geometrically thin and optically thick disk (Shakura \& Sunyaev 1973).
The gravitational power dissipated in an accretion disk
through a viscous process per unit surface area is
\begin{equation}
Q_{\rm vis}^+={\frac {3GM_{\rm BH}\dot{M}}{8\pi R^3}}
\left[1-\left({\frac {3R_{\rm s}}{R}}\right)^{1/2}\right],
\label{q_vis}
\end{equation}
where $R_{\rm S}=2GM_{\rm BH}/c^2$ is the Schwarzschild radius.

The equation of state of the accretion disk is
\begin{equation}
P_{\rm t, d}=P_{\rm g,d}+P_{\rm r,d}+P_{\rm B}=(P_{\rm g,d}+P_{\rm r,d})(1+1/\beta_{\rm 0})=({\frac {\rho_{\rm d} kT_{\rm d}}{\mu
m_{\rm p}}}+{\frac {1}{3}}aT_{\rm d}^4)(1+1/\beta_{\rm 0}), \label{gas state}
\end{equation}
where magnetic pressure $P_{\rm B}=B^2/8\pi$ is characterized by the magnetic parameter $\beta_{\rm 0}$, i.e., the ratio of the sum of gas pressure and radiation pressure to magnetic pressure.

We assume that the magnetic field is continually generated in the disk by dynamo action. Because of the buoyancy instability,
the magnetic flux loops can emerge into the corona and reconnect with other loops. In this process, a fraction ($f$)
of the gravitational energy stored in
the magnetic field is transferred into the corona, i.e.,
\begin{equation}
Q_{\rm c}^+= \frac{B^2}{4\pi}V_{\rm A}=fQ_{\rm vis}^+,
\label{q_frac}
\end{equation}
where  Alfv\'{e}n speed
$V_{\rm A}=\sqrt{2P_{\rm B}/\mu m_{\rm H}n_{\rm c}}$ and 
$n_{\rm c}$ is the number density of electrons in the corona.
Here, we take the mean molecular weight $\mu$ to be 0.5, which is case of the chemical composition of pure hydrogen.

 As shown in former works, the energy equilibrium of the disk is determined by the accretion energy released in the disk and irradiation by the corona (Haardt \& Maraschi 1991, 1993, Cao 2009). In our present work, it is assumed that some parts of the seed photons from the disk are upward scattered as the emergent spectrum, and parts of them are scattered backward. The backward photons are reprocessed  in the disk surface layer and emitted as blackbody. This means that the irradiation by the corona can only affect the disk blackbody temperature rather than the internal structure of the disk (Tuchman et al. 1990; Liu et al. 2003). Thus, the energy equation for the cold disk is,
\begin{equation}
Q_{\rm vis}^{+}(1-f)=\frac {8\sigma T_{\rm d}^4}{3\tau},
\label{disk_energy}
\end{equation}
where $\tau=2.0\rho_{\rm d} H_{\rm d} \kappa$ is the optical
depth in the vertical direction of the disk. The height of the disk is
$H_{\rm d}= c_{\rm s}/{\Omega}=\sqrt{P_{\rm t,d}/\rho_{\rm d}}/{\Omega}$. The opacity $\kappa$ is contributed by the scattering opacity $\kappa_{\rm es}$ and free-free opacity $\kappa_{\rm ff}$. We take $\kappa_{\rm es}=0.4\,\rm ~{cm^{2}~g^{-1}}$ for the chemical composition of pure hydrogen and
$\kappa_{\rm ff}=6.4\times10^{22}\rho_{\rm d}T_{\rm d}^{-7/2}\,\rm ~{cm^{2}~g^{-1}}$.

A fraction ($f$) of angular momentum is carried into the corona along
the magnetic loops, and the remaining fraction ($1-f$) of it
is maintained in the disk, which is shown in the following equation:
\begin{equation}
\dot{M}\Omega\left[1-\left({\frac {3R_{\rm
s}}{R}}\right)^{1/2}\right](1-f)=4\pi H_{\rm d}\tau_{\rm r\varphi},
\label{angular momentum}
\end{equation}
where $\tau_{\rm r\varphi}=\alpha P_{\rm t,d}$ is the viscosity stress.

The energy transferred from the disk is released in the corona and eventually radiated away mainly via inverse
Compton scattering. The density of the corona is
determined by the energy balance between the downward thermal conduction and the mass evaporation in the chromospheric layer.
So the energy equations for the corona can be summarized as
\begin{equation}
\label{corona energy}
\frac{B^2}{4\pi}V_{\rm A}\approx \frac{4kT_{e}}
{m_{e}c^2}\tau^*cU_{\rm rad},
\end{equation}

\begin{equation}\label{e:evap}
{k_0T_{ e} ^{7\over 2}\over \ell_{\rm c}}\approx
{\gamma\over \gamma-1} n_{\rm c} k T_{e}
\left(\frac{kT_{e }} {\mu m_{\rm H}}\right)^{2}.
\end{equation}
In equation (\ref{corona energy}), the energy density of soft  photons is
$U_{\rm rad}=\frac{2}{c}Q_{\rm vis}^{+}(1-f)+0.4\lambda_{\rm u}\frac{B^2}{8\pi}$,
which includes both the intrinsic disk radiation and the
reprocessed radiation of backward Compton emission (with
albedo always being assumed to be zero in our calculations).
In equation(\ref{corona energy}), $\tau^*\equiv\lambda_{\rm t}\tau=\lambda_{\rm t}n_{\rm c}\sigma_{\rm T}\ell_{\rm c}$
is the effective optical depth, with an initial value of $\lambda_{\rm t}=1$. In our model,
we also set $\ell_{\rm c}=10R_{\rm S}$
as did in previous works (e.g.,  Liu et al. 2002b, 2003; Qiao \& Liu 2015), since it is found that the emergent spectrum is weakly dependent on $\ell_{\rm c}$.

Given the values of black hole mass $M_{\rm BH}$, radius $R$, mass accretion
rate $\dot{m}$, viscous coefficient $\alpha$, magnetic parameter $\beta_{\rm 0}$,
and initial parameters $\lambda_{\rm t}=1.0$ and $\lambda_{\rm u}=1.0$,
we solve equations (\ref{q_vis})-(\ref{e:evap}) numerically and obtain
the radially dependent disk temperature $T_{\rm d}$,
disk density $\rho_{\rm d}$, the temperature $T_{e}$, and number
density of electrons $n_{c}$ in the corona.

With the parameters of the structure of the disk corona, we derive the spectrum of the
model through Monte Carlo simulation. This method is essentially the same
as that described by Pozdniakov et al. (1977), and the detail
process was shown in detail in Liu et al. (2003). In our calculation, we consider both the
intrinsic disk  photons and reprocessed photons by the disk  as the seed photons to be scatterred in
the corona. In order to get a self-consistent solution, we need to check
whether the upward luminosity from the corona $L_{\rm up}$ is approximately equal to $L_{G}$ (here $L_{G}$ is the liberated rate of total gravitational energy), and the ratio of the downward luminosity to the soft luminosity $L_{\rm down}/L_{\rm soft}$
is approximately equal to $\lambda_{\rm u}'$ (here $\lambda_{\rm u}'$ is the ratio of the energy of reprocessed photons to the total soft photon energy in structure calculation). If yes, we find the right $\lambda_{\rm t}$
and $\lambda_{\rm u}$ for the consistent corona structure. If no,
we set new $\lambda_{{\rm u},n+1}=(L_{{\rm down},n}/L_{{\rm  soft},n})/
\lambda_{{\rm u},n}'\times\lambda_{{\rm u},n}$ and $\lambda_{{\rm t},n+1}=
(L_{ {\rm up},n}/L_{\rm G})\times\lambda_{{\rm{t}},n}$ and then repeat
the structure calculation and Monte Carlo simulation until the consistent
 conditions $L_{\rm up}\approx L_{G}$ and
$L_{{\rm down},n}/L_{ {\rm soft},n}\approx \lambda_{{\rm u},n}'$ are fulfilled.

\section{RESULTS}
\subsection{Structure of the Disk and Corona}\label{structure}

Given the black hole mass $M_{\rm BH}=10^8M_\odot$ and viscous coefficient $\alpha=0.3$ as others (as in, e.g., King et al. 2013), we numerically solve equations (\ref{q_vis})-(\ref{e:evap}) to obtain the radial structure of the disk corona. Since we aim to investigate the X-ray spectrum properties of luminous AGNs, the accretion rate $\dot{m} \geq 0.03$ in this work.

\begin{figure}
\centering
\includegraphics[width=5.2in,height=3.0in]
{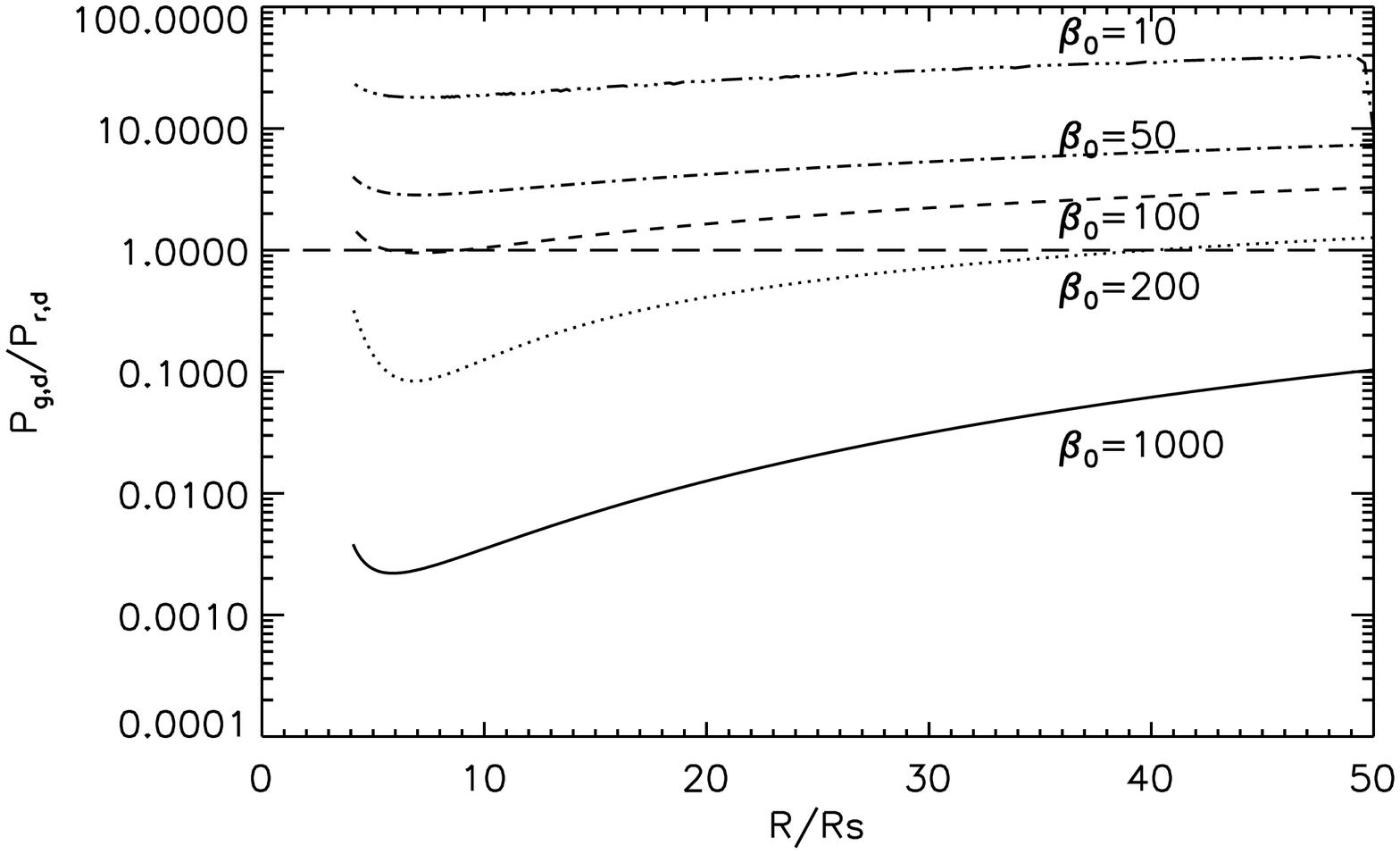}
\includegraphics[width=5.2in,height=3.0in]
{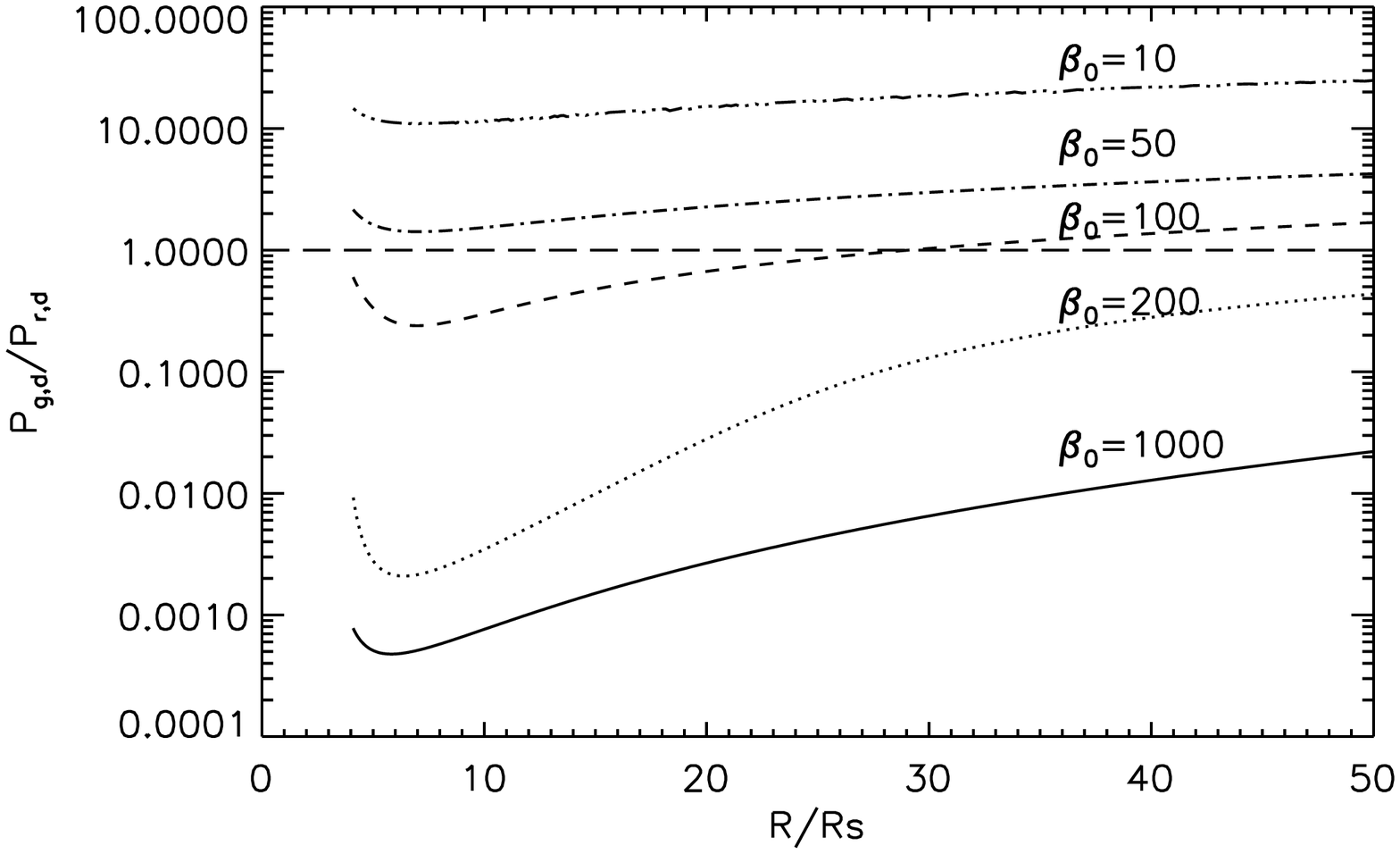}

\caption{Radial distribution of the ratio of gas pressure to radiation pressure ($P_{\rm g,d}/P_{\rm r,d}$) in the disk
 for various $\beta_{\rm 0}$ and two values of accretion rate $\dot{m}=0.05$ (upper panel)
 and 0.1 (lower panel). Solid line: $\beta_{\rm 0}=1000$; the dotted line:
 $\beta_{\rm 0}=200$; dashed lines: $\beta_{\rm 0}=100$; dot-dashed line: $\beta_{\rm 0}=50$ and the triple-dot-dashed line: $\beta_{\rm 0}=10$. The long-dashed line
 denotes that $P_{\rm g,d}/P_{\rm r,d}=1.0$.}
\label{Fig:mdot0.1-p-0526}
\end{figure}

  Numerical calculation shows that the radial structure of the disk is sensitively affected by the magnetic field. We show the radial distribution of the ratio of gas pressure to radiation pressure ($P_{\rm g,d}/P_{\rm r,d}$) in the disk for accretion rate $\dot{m}=0.05$ and $\dot{m}=0.1$ with different $\beta_{\rm 0}$ in Fig.~\ref{Fig:mdot0.1-p-0526}. This shows that the ratio of $P_{\rm g,d}/P_{\rm r,d}$ becomes larger as the $\beta_{\rm 0}$ decreases. When $\beta_{\rm 0}$ is 1000, $P_{\rm g,d}/P_{\rm r,d}$ is always less than 1 in all the regions of $R < 50~R_{\rm s}$ for both accretion rate, which means that the disk is absolutely dominated by the radiation pressure for this weak magnetic field. For a larger magnetic field with $\beta_{\rm 0}=200$, $P_{\rm g,d}/P_{\rm r,d}$ is larger than 1 in the range of $R > 40~R_{\rm s}$ for $\dot{m}=0.05$. However, for $\dot{m}=0.1$, the ratio $P_{\rm g,d}/P_{\rm r,d}$ is still less than 1 in the disk. When $\beta_{\rm 0}=100$, the region dominated by $P_{\rm g,d}$ is expanded to $R > 10~R_{\rm s}$ for $\dot{m}=0.05$ and to $R > 30~R_{\rm s}$ for $\dot{m}=0.1$. For $\beta_{\rm 0}\leq 50$, $P_{\rm g,d}/P_{\rm r,d}$ is larger than 1 in the region of $R < 50~R_{\rm s}$ for both accretion rates, which means that the disk is absolutely dominated by of $P_{\rm g,d}$.

\begin{figure}

 \centering
 \includegraphics[width=5.2in,height=3.0in]
 {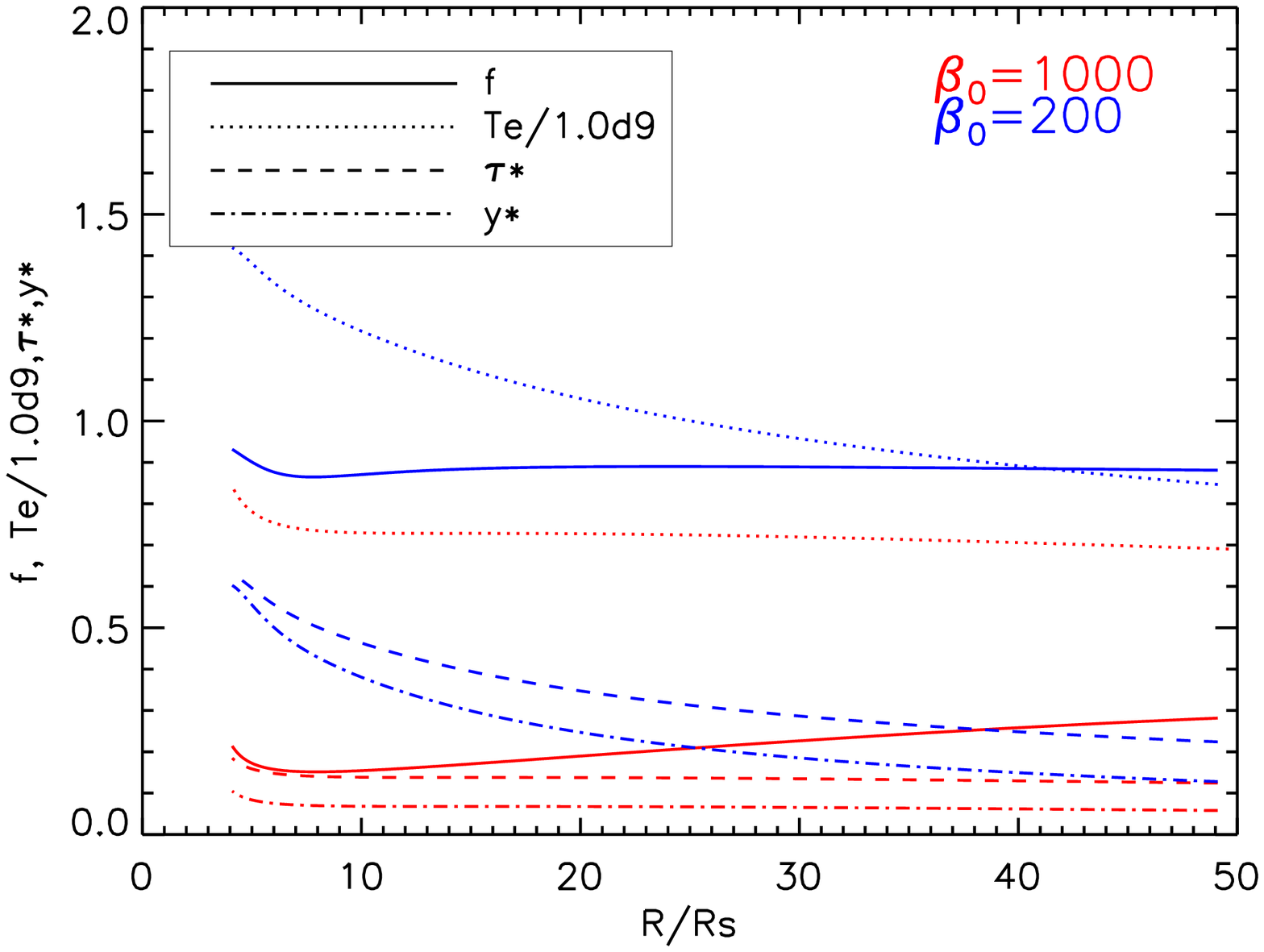}
 \includegraphics[width=5.2in,height=3.0in]
 {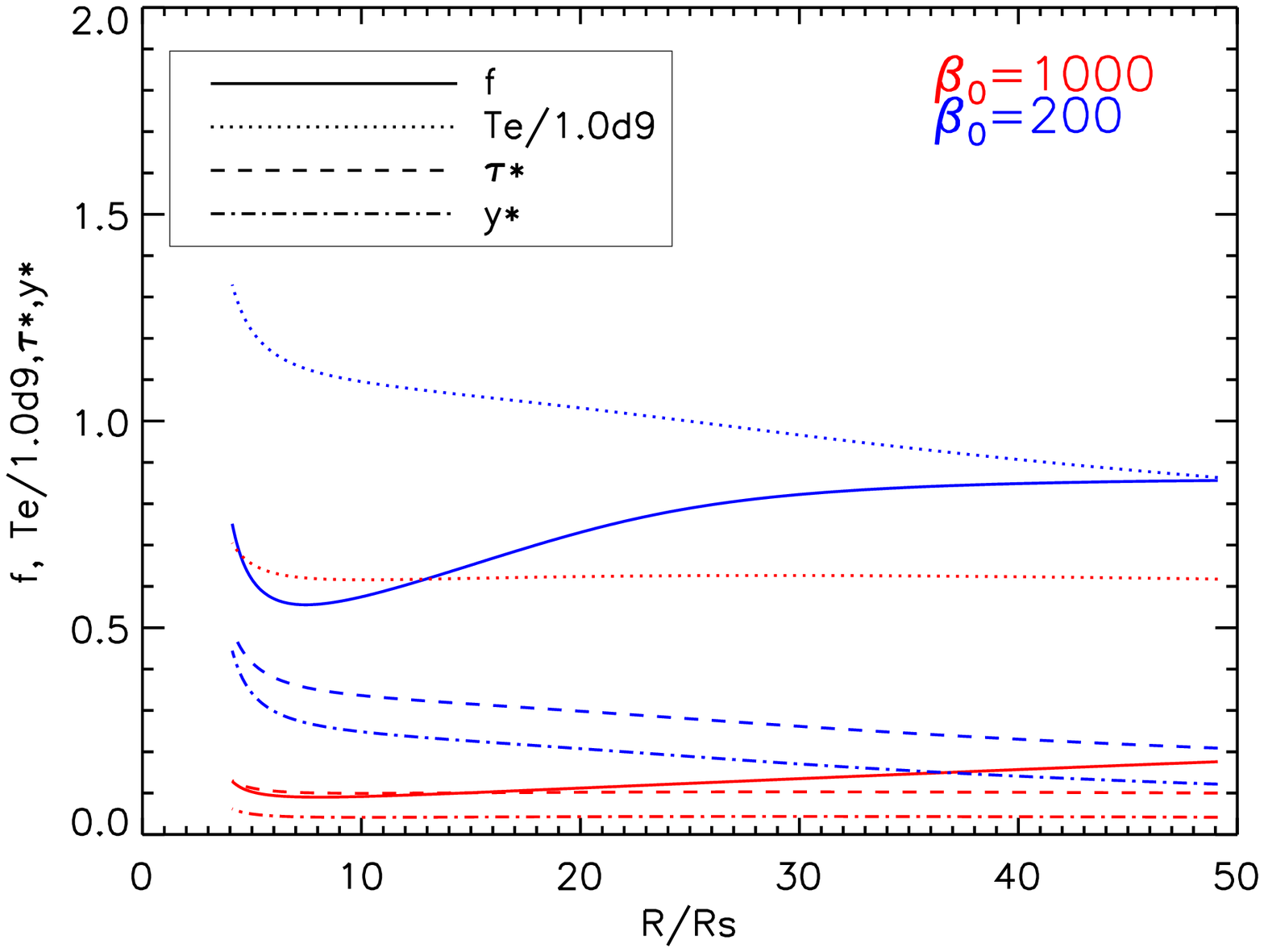}

 \caption{Radial distribution of parameters in the corona with different $\beta_{\rm 0}$ (red lines: 1000; blue lines: 200) for $\dot{m}=0.05$ (upper)
  and $\dot{m}=0.1$ (lower). All the parameters increase with a decrease
  of $\beta_{\rm 0}$, which means that the corona becomes stronger for larger
  magnetic field strength.}
 \label{Fig:mdot0.1-structure-0526}
 \end{figure}

 In order to investigate the features of the corona under the effect of magnetic reconnection heating, we show the radial distribution
 of the energy fraction $f$, electron temperature $T_e/10^9$, effective optical depth $\tau^*$ and effective Compton $y$-parameter
 ($y^*=4kT_e/m_ec^2$) in the corona with various $\beta_{\rm 0}$ for
 $\dot{m}=0.05$ and $\dot{m}=0.1$,  respectively in Fig.\ref{Fig:mdot0.1-structure-0526}. For $\dot{m}=0.05$, the energy fraction $f$ is dramatically increased from about 0.2 to 0.9 as $\beta_{\rm 0}$ decreases from 1000 to 200. Similar changes can be found for $\dot{m} = 0.1$ in the lower panel of Fig.\ref{Fig:mdot0.1-structure-0526}, which shows that $f$ increases from 0.1 to larger than 0.5 with the decrease of $\beta_{\rm 0}$ from 1000 to 200. For $\beta_{\rm 0}=1000$, the energy fraction is less,  most of the gravitational energy is liberated in the disk, and the corona is weak with a temperature of about $10^8\rm K$ (shown by red dotted lines in Fig.2).For $\beta_{\rm 0}=200$, the electron is effectively heated to about $10^9$~K (blue dotted lines in Fig.2). The effective optical depth $\tau^*$ is also increased as the result of efficient evaporation at the interface between the corona and disk. Since the temperature and the effective optical depth increase with the stronger magnetic field, the effective $y$-parameter $y^*$ also becomes larger at less $\beta_{\rm 0}$ as shown in Fig.\ref{Fig:mdot0.1-structure-0526}.

\begin{figure}
\centering

\centering
\includegraphics[width=5.2in,height=3.0in]
{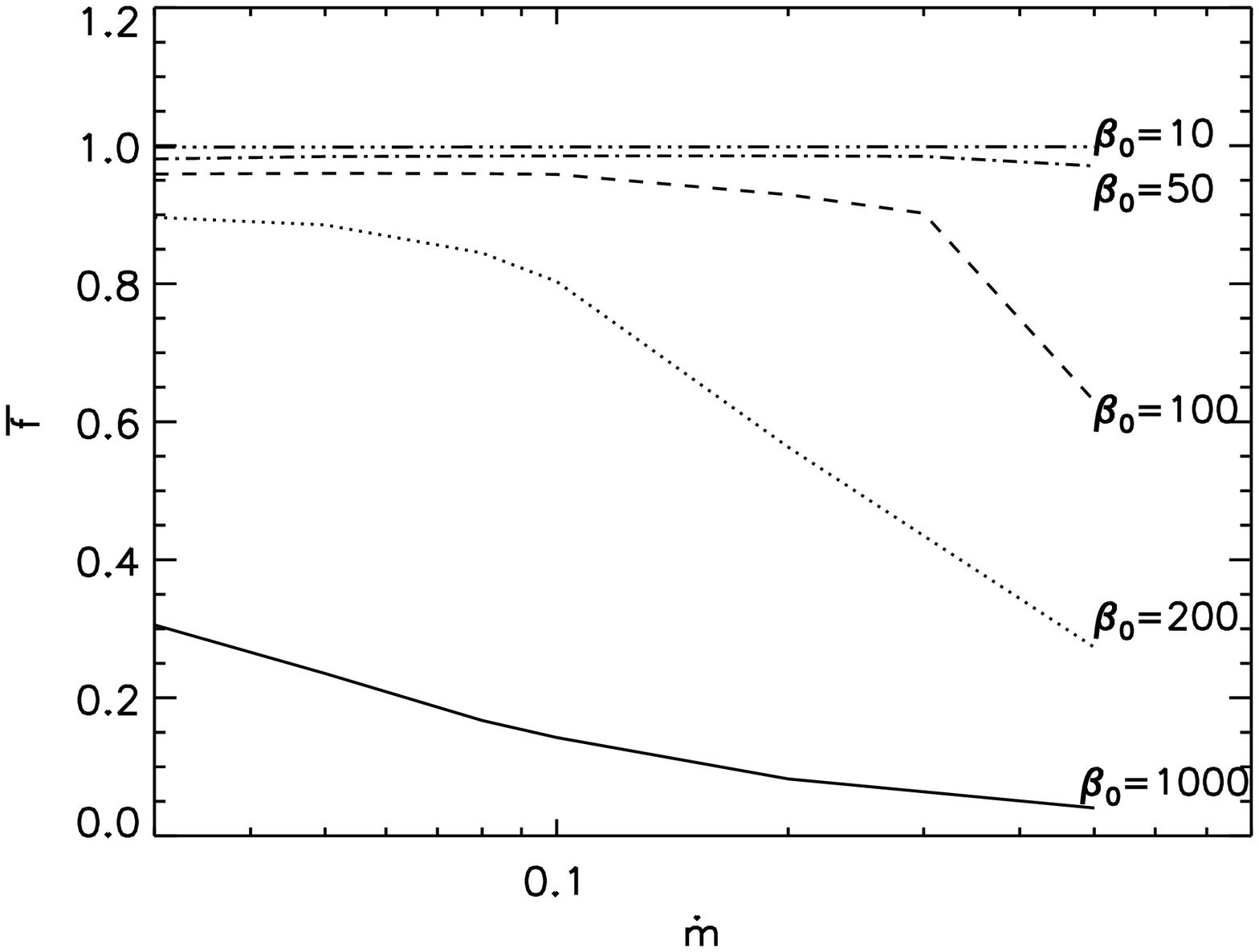}
\includegraphics[width=5.2in,height=3.0in]
{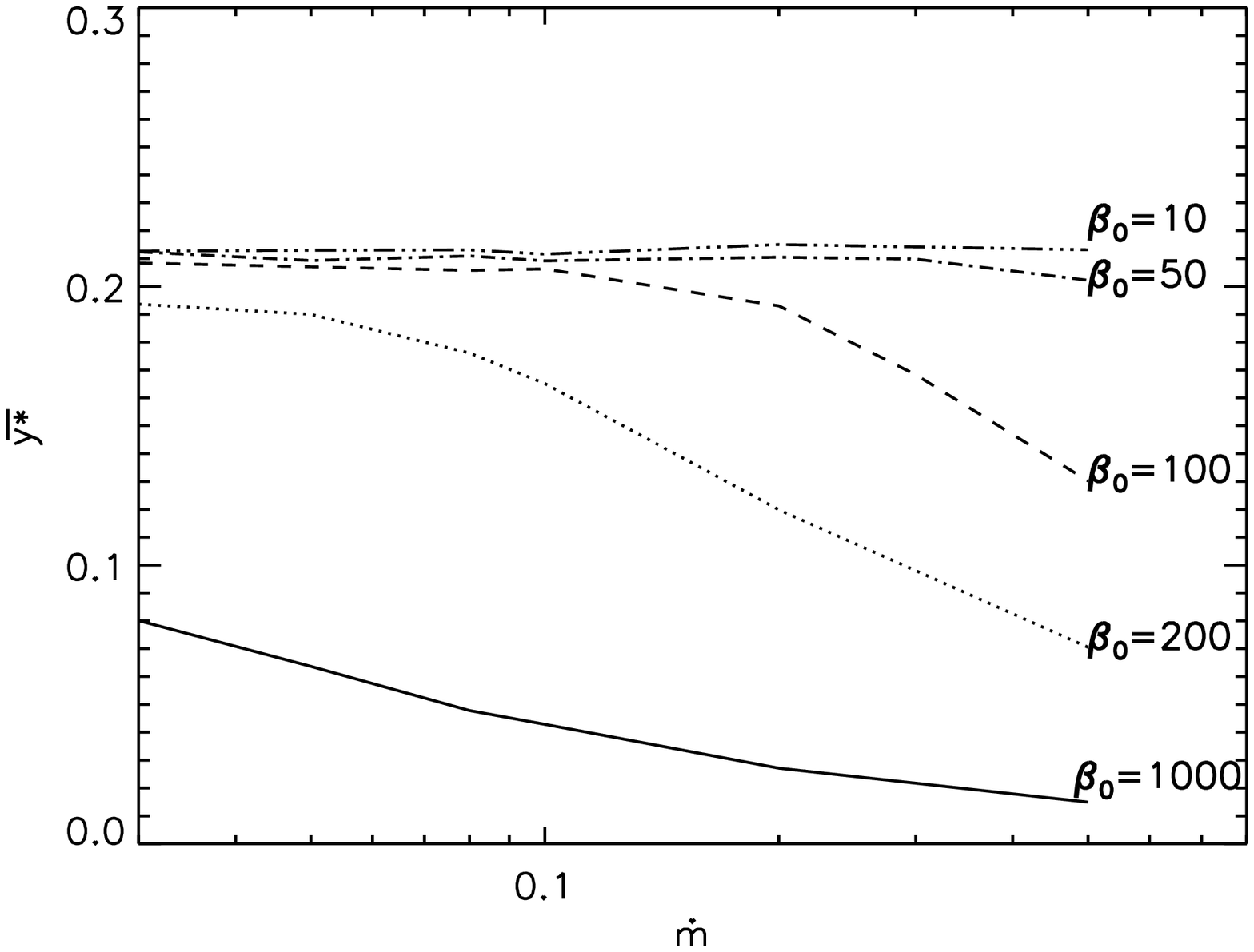}
 \caption{Averaged energy fraction
$\overline{f}$ (upper panel)
and averaged effective Compton $y$-parameter $\overline{y^*}$ (lower panel) versus accretion rate for different $\beta_{\rm 0}$. The denotations of different line styles are the same as those in Fig.~1. Both of these two parameters decrease with the increase of accretion rate for $\beta_{\rm 0}>100$ and are the same for all the accretion rates for $\beta_{\rm 0}=50 (10)$.}
\label{Fig:average_f_y}
\end{figure}

 We also show the relation between accretion rate and averaged energy fraction $ \overline {f}$ ($=\frac{\int f\times2\pi RdR}{\int2\pi RdR}$)
 and averaged effective Compton $y$-parameter $ \overline {y^*}$ ($ = \frac{\int y^*\times2\pi RdR}{\int2\pi RdR}$) for the same $\beta_{\rm 0}$ in Fig. \ref{Fig:average_f_y}.
 When the corona is heated by weak magnetic reconnection with $\beta_{\rm 0}=1000$, both $ \overline {f}$ and $ \overline {y^*}$ are low and they decrease as accretion rate increases. For the same accretion rate, both $ \overline {f}$ and $ \overline {y^*}$ increase as magnetic field increases. This means that more energy is carried into the corona by stronger magnetic field, which also results in stronger emission in the corona. However, when $\beta_{\rm 0}=10$, $\overline {f}$ is nearly equal to 1 and $\overline {y^*}$ is also constant, 0.2, for different accretion rates. This implies that almost all gravitational energy is carried into the corona and the most powerful corona is formed when $\beta_{\rm 0}=10$. Theses results are similar to that the case of disks dominated by gas pressure in Liu et al.(2003).

\subsection{Spectrum from the Disk and Corona}
\begin{figure}
\centering
\includegraphics[width=5.2in,height=6.0in]{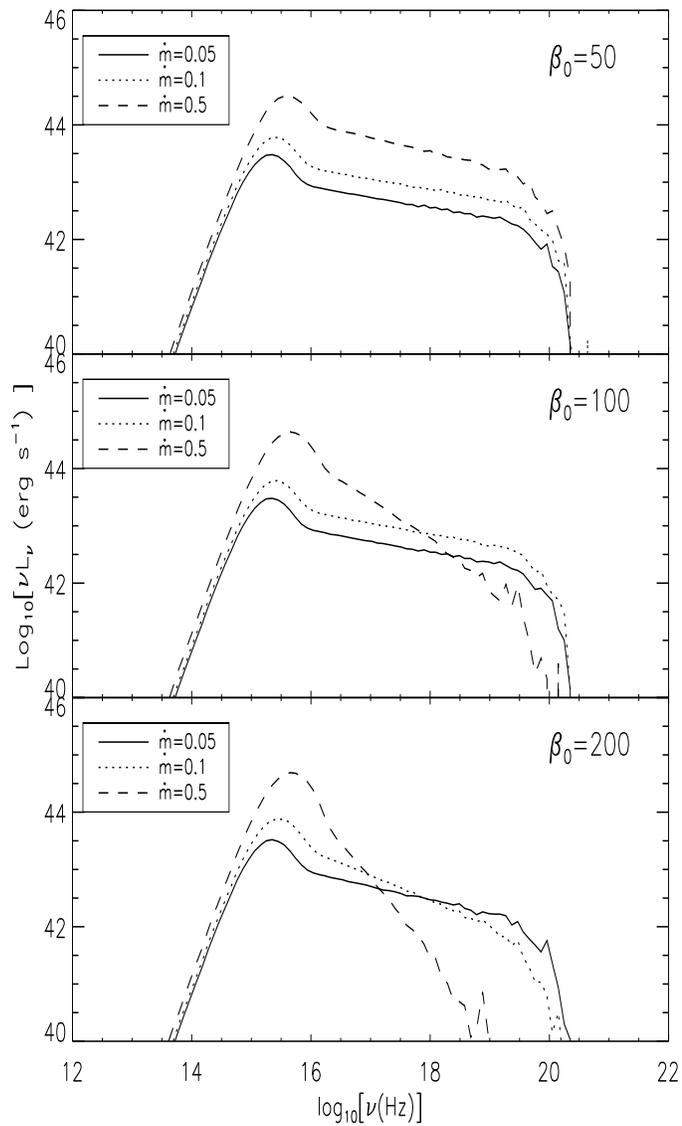}
  \vspace{12mm}
 \caption{Spectra of the disc corona
  with different $\beta_{\rm 0}$
  (upper: 50; middle: 100; lower: 200). The
  accretion rates in each figure are 0.05 (solid lines),
   0.1 (dotted lines), and 0.5 (dashed lines) respectively.}
\label{Fig:data-spect-beta-10-200-1000}
\end{figure}

  With the radial structure, the spectrum of the disk corona can be derived by Monte Carlo simulation.
    We plot the spectrum
    with different $\beta_{\rm 0}$ (200, 100, 50) for three different accretion rates $\dot{m}$ (0.05, 0.1, 0.5) in Fig. \ref{Fig:data-spect-beta-10-200-1000}.
    For a magnetic filed with $\beta_{\rm 0}=200$, the hard X-ray is weaker for higher accretion rates. This shows that the spectrum of $\dot{m} = 0.5$ is mainly contributed by the blackbody radiation from the accretion disk. When $\beta_{\rm 0} = 100$, both the spectra of $\dot{m} = 0.05$ and $\dot{m} = 0.1$ are slightly harder than that of $\beta_{\rm 0} = 200$ and seem similar, whereas the hard X-ray spectrum of $\dot{m} = 0.5$ dramatically becomes harder than that in the case of $\beta_{\rm 0} = 200$. As $\beta_{\rm 0}$ decreases to 50, the X-ray spectra for these three accretion rates are similar, as shown in the upper panel of Fig. \ref{Fig:data-spect-beta-10-200-1000}.  As we mentioned in the previous section, the averaged effective Compton $y$-parameter $\overline{y^*}$ is independent of the accretion rate when $\beta_{\rm 0}=50$, which leads to the hard X-ray spectrum hardly changing with the accretion rate. The spectrum is similar to the hard spectrum shown in Fig.5 in Liu et al. (2003).

    \subsection{Application in Luminous AGNs}

\begin{figure}

\centering
\includegraphics[width=5.2in,height=3.00in]{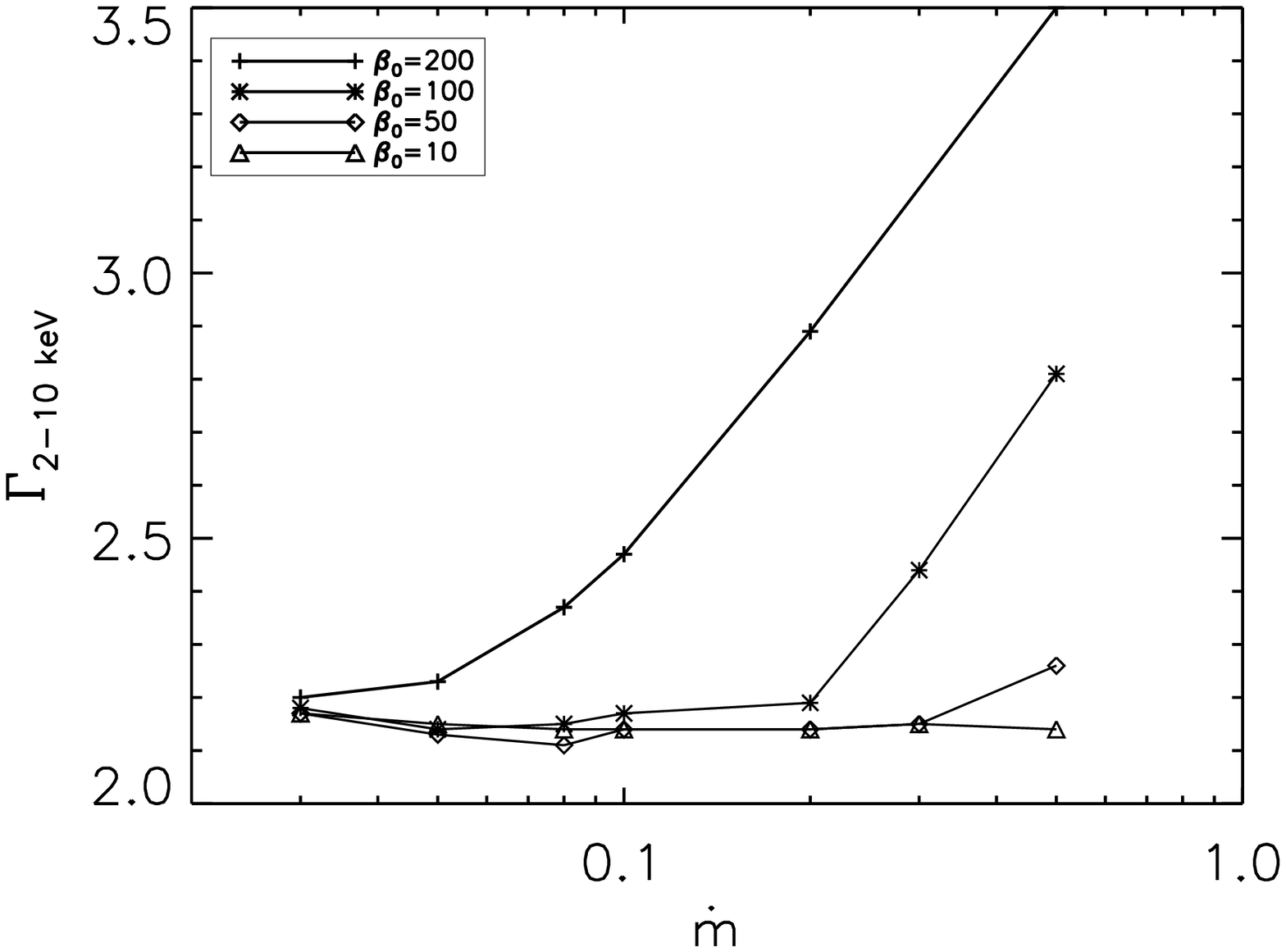}

\includegraphics[width=5.2in,height=3.0in]{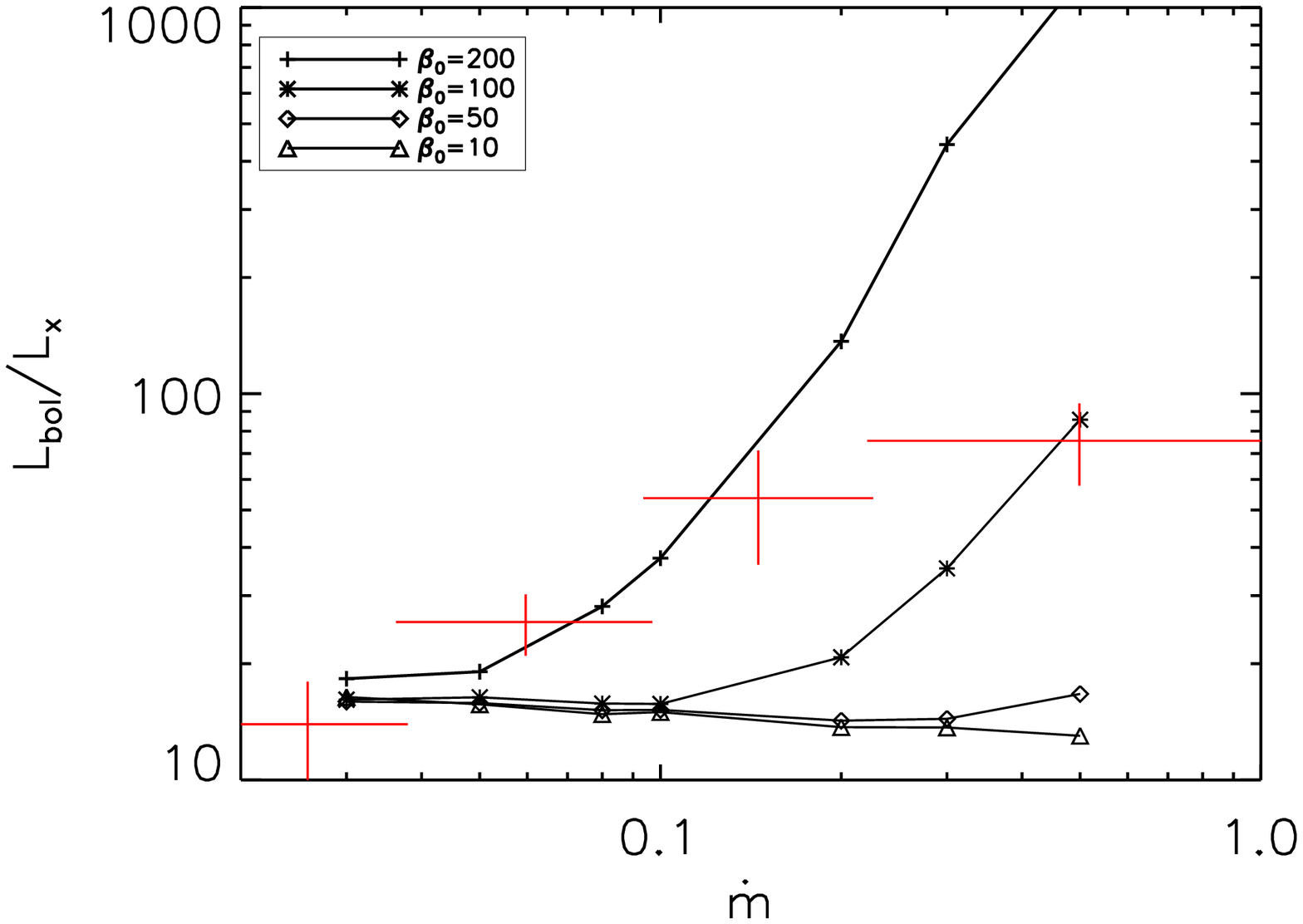}

 \caption{Upper panel: the hard X-ray photon index
 ($\Gamma_{\rm 2-10~keV}$) as a function of accretion rate for different $\beta_{\rm 0}$. Lower panel: the hard X-ray bolometric correction factor $L_{\rm bol}/L_{\rm x}$ as a function of accretion rate for different $\beta_{\rm 0}$. The observed data taken from Vasudevan \& Fabian(2009) are shown by
   the big red crosses. The lines with $\beta_{\rm 0}=100\,, 200$ fit the data well. However, the lines with $\beta_{\rm 0}=50\,, 10$ are not consistent with the observed result.} 
\label{Fig:mdot-lbol-lx}
\end{figure}

 In luminous AGNs, the observational results show that both $\Gamma_{\rm 2-10~keV}$ and $L_{\rm bol}/L_{\rm x}$ increase with accretion rate. For comparison, we plot the relation between $\Gamma_{\rm 2-10~keV}$ (and $L_{\rm bol}/L_{\rm x}$) and accretion rate for different $\beta_{\rm 0}$ in Fig.\ref{Fig:mdot-lbol-lx}. For $\beta_{\rm 0}=200$, $\Gamma_{\rm 2-10~keV}$ is about 2.2 at accretion rate $\dot{m} = 0.03$ and it increases to about 3.5 at $\dot{m} = 0.5$. Correspondingly, $L_{\rm bol}/L_{\rm x}$ increases from 19 to larger than $10^4$, with the accretion rate increasing from 0.03 to 0.5. When $\beta_{\rm 0}=100$, $\Gamma_{\rm 2-10~keV}$ is about
  2.17 at $\dot{m}<0.1$, and it also increases to 2.8 at $\dot{m} = 0.5$. However, for 
a larger magnetic field (shown by lines with $\beta_{\rm 0}=50$, 10), $\Gamma_{\rm 2-10~keV}$ is about 2.17 for all the accretion rates, which is consistent with the results in previous work s(Liu et al. 2003, Cao 2009, Kawabata \& Mineshige 2010). Correspondingly, $L_{\rm bol}/L_{\rm x}$ is also constant at 16.

  In order to compare with the observed results in luminous AGNs, in the lower panel of Fig.\ref{Fig:mdot-lbol-lx}, the observational data (represented by big red crosses) are also plotted. These data are selected from Vasudevan \& Fabian (2009), which are the binned observational data with radio-loud objects and low X-ray flux objects removed. Drawn from the plot, for $\beta_{\rm 0} = 200$, the model fits the observational data well at $0.05 < \dot{m} < 0.2$. It predicts a relatively higher value of $L_{\rm bol}/L_{\rm x}$ than that observed at $\dot{m} > 0.2$, which means that the corona is still energy inadequate for the X-ray emission. We also find that $\beta_{\rm 0} = 100$ is suitable for the observation at $\dot{m} > 0.2$, whereas the model with $\beta_{\rm 0} = 100$ predicts larger X-ray emission, i.e., less $L_{\rm bol}/L_{\rm x}$ than the observed for $\dot{m} < 0.2$. However, $L_{\rm bol}/L_{\rm x}$ is always constant with $\beta_{\rm 0} = 50$ and $\beta_{\rm 0} = 10$ for different $\dot{m}$, which is also not consistent with the observational results. In general, we suggest that $\beta_{\rm 0}$ should be about 100-200 with $\alpha\,=\,0.3$ in accretion disk for hard X-ray emission in luminous AGNs. Combined with the averaged energy fraction for $\beta_{\rm 0}=100-200$ shown in the upper panel of Fig.3, we suggest that larger than about $30\%$ of the total gravitational energy is needed to be carried into the corona through magnetic reconnection in luminous AGNs.

\section{DISCUSSION}

\subsection{The Relation between $\alpha$ and $\beta$}
Simulation works show that the viscous coefficient $\alpha$ increases with the decreasing of $\beta$ and their product remains nearly constant (e.g.,  Blackman et al. 2008; Guan et al 2009; Hirose et al. 2009; Sorathia et al. 2012). Blackman et al.(2008) showed that the product is ~0.5. However, the values of $\alpha$ and $\beta$ vary from one simulation to the other (Yuan \& Narayan 2014). Besides, the value of $\alpha$ in simulations also deviates from the value constrained by observation (King et al. 2013, Liu \& Taam 2013).

\begin{figure}
\centering
\includegraphics[width=5.2in,height=4.0in]{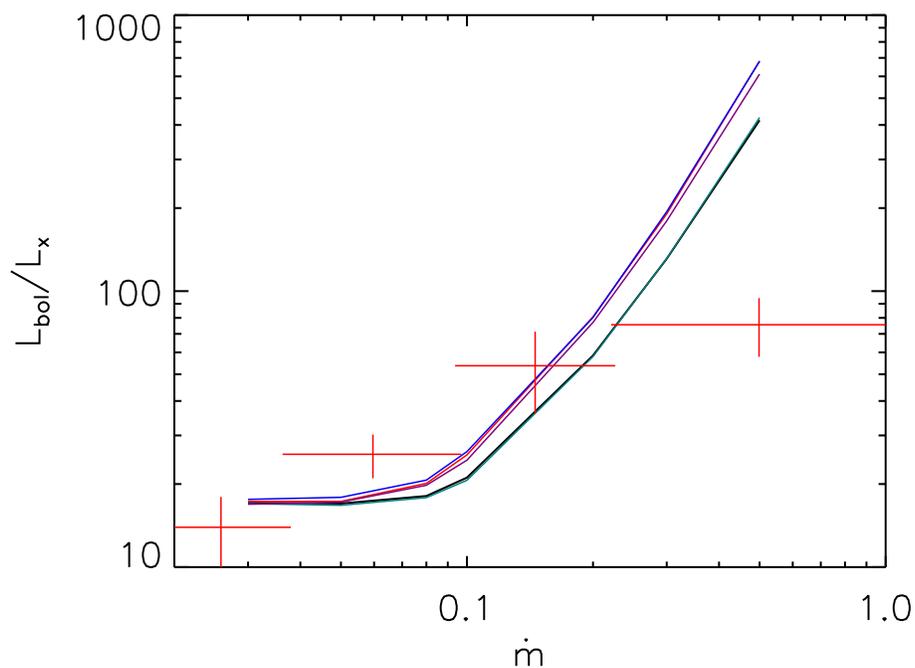}
 \caption{Relation between $L_{\rm bol}/L_{\rm x}$
 and accretion rate $\dot{m}$ derived from the model for different $\alpha$ and $\beta_{\rm 0}$. Red line is for $\alpha\,=\,0.1$ and $\beta_{\rm 0}=500$; black line is for $\alpha\,=\,0.3$ and $\beta_{\rm 0}=150$; blue line is for $\alpha\,=\,0.5$ and $\beta_{\rm 0}=100$; Purple line is for $\alpha\,=\,0.7$ and $\beta_{\rm 0}=70$; Dark cyan line is for $\alpha=0.9$ and $\beta_{\rm 0}=50$. These lines predict similar correlation between $L_{\rm bol}/L_{\rm x}$ and $\dot{m}$, which are roughly consistent with the observed result. The data taken from Vasudevan \& Fabian (2009) are also shown by the big red crosses.} 
\label{Fig:mdot-lbol-lx-alpha-obs}
\end{figure}

The effect of magnetic field on the accretion flow is commonly investigated by changing the value of $\beta$ for certain $\alpha$ (Qian et al. 2007; Li \& Begelman 2014). In our present work, we aim to overcome the inadequate energy in the corona of luminous AGNs and suppose that the corona is formed through the magnetic reconnection process. Given $\alpha\,=\,0.3$ as in the observational result, we change the value of $\beta_{\rm 0}$ to find the proper magnetic field in accretion flow to fit the observed X-ray emission in luminous AGNs. It is found that the thin disk changes from being radiation pressure dominated to being gas pressure dominated as the magnetic field increases. Correspondingly, the spectrum becomes harder for larger magnetic field. We also find an appropriate value of $\beta_{\rm 0}$ for different $\alpha$. The relations between $L_{\rm bol}/L_{\rm x}$ and $\dot{m}$ for different $\alpha$ and $\beta_{\rm 0}$ in our model, which fit well with the observed data, are shown in Fig.\ref{Fig:mdot-lbol-lx-alpha-obs}. It is found that for less $\alpha\,=\,0.1$ (red lines), $\beta_{\rm 0}$ is about 500, whereas $\beta_{\rm 0}$ decreases to about 50 for $\alpha\,=\,0.9$ (dark cyan lines). This means that larger $\alpha$ requires stronger magnetic field, which is roughly consistent with the MHD simulation results.

\subsection{Is the Corona Strongly Magnetic Field Supported?}
In our present work, we suppose that the magnetic field is generated at the midplane of the disk. The magnetic loops erupt into the corona because of the buoyancy instability and reconnect with other loops. Therefore, the magnetic flux is released to heat electrons in the corona. Similar to the structure of the solar corona, the low-$\beta$ corona is also dynamically controlled by the magnetic field whose footpoints are embedded in buoyantly unstable, $\beta\sim 1$ plasma (Shibata et al. 1989, Di Matteo 1998). Miller \& Stone (2000) found that the strongly magnetized and stable corona with $\beta\leq0.1$ can be formed through turbulence in the disk with the initial weak magnetic field. $\beta$ varies by about 3 orders of magnitude from the corona to the disk midplane. We find that $\beta_{\rm 0} \sim 50-500$ with $\alpha\,=\,0.9-0.1$ is suitable for the X-ray emission in luminous AGNs. When $\beta_{\rm 0}\sim 50-500$, $\beta$ approaches about $50-500$ in the disk because of the disk being dominated by gas pressure. According to the vertical distribution of magnetic field found in Miller \& Stone (2000), we can roughly estimate that $\beta \sim 0.05-0.5$ in the corona. This might indicate that the magnetic parameter $\beta$ in the corona of luminous AGNs is less than that in LLAGNs found by Qiao \& Liu (2013). They fitted the X-ray observational results in LLAGNs within the framework of the disk evaporation model and found that the magnetic field in the corona should be weak, i.e., $\beta\,=\,4-19$.

 \subsection{Comparing with Other Hybrid Disk-corona Models}
 As shown by the corona energy equation (\ref{corona energy}), it seems that the corona is
sensitively affected by the amount of the energy carried by the magnetic reconnection. In other words, the value of magnetic pressure $P_{\rm B}$ determines the structure and spectral features of the disk corona.

 In Cao (2009) and You et al. (2012), they
constructed a disk-corona model with three different types of magnetic stress tensor. The corona is two-temperature. The temperature of the ions is set to be 0.9 of the virial temperature and electrons are heated through Coulomb collision with ions. It was tested that the magnetic stress tensor as $\tau_{\rm r\varphi}=P_{\rm B}=\alpha P_{\rm t}$ always leads to constant $L_{\rm bol}/L_{\rm x}$ for different accretion rates. At the lowest accretion rate, the photon indices stay between 2.0 and 2.2. In our work, we take $\beta_{\rm 0}=(P_{\rm r,d}+P_{\rm g,d})/P_{\rm B}$ and assume that the electron is directly heated by the  energy released in the process of magnetic reconnection. It seems that there is a relation between $\alpha$ (in that work) and $\beta_{\rm 0}$,
i.e., $\beta_{\rm 0}=\frac{1}{\alpha}-1$. For $\alpha\,=\,0.3$, $\beta_{\rm 0}$ should be 2.33 in our work. We can compare the value of $\Gamma_{\rm 2-10 ~keV}$ and $L_{\rm bol}/L_{\rm x}$ for $\beta_{\rm 0}=10$ (the same magnitude order as 2.33) with the results found in Cao (2009). For $\beta_{\rm 0}=10$, nearly all of the total gravitational energy is carried into the corona by the efficiently magnetic reconnection process, and the strongest corona is formed. In fact, it is also found that $\Gamma_{\rm 2-10 ~keV}$ and $L_{\rm bol}/L_{\rm 2-10~keV}$ hardly change with accretion rate, which is similar to the results found in Cao (2009). However, when $\beta_{\rm 0} >10$, the magnetic field is weaker than the case of $\beta_{\rm 0}=10$ for the same accretion rate. Thus, the total spectrum of the disk corona becomes softer with weaker magnetic field.

We note that the hard X-ray photon index $\Gamma_{\rm 2-10~keV}$ is always larger than 2.0 at the lowest accretion rate considered in our model. However, there are many luminous AGNs whose $\Gamma_{\rm 2-10~keV} < 2.0$ (Yang et al. 2015). These sources might accrete through the clumpy two-phase accretion flow or accretion flows with coronas condensing into disks (Yang et al.2015; Liu et al. 2007; Liu \& Taam 2009; Qiao \& Liu 2013; Liu et al. 2015). The main reason might be the difference in the soft photon field for inverse Compton scattering in the hot corona. In our present work, both the reprocessed corona photons and the intrinsic disk photons are included for the inverse Compton scattering. Even though nearly all of the gravitational energy is carried into the corona, the continuous accretion disk absorbs the backward inverse Compton emission, and the gas in the chromosphere is efficiently heated, which results in sufficient soft photons for the Compton cooling in the corona.  While for the clumpy cold gas or the condensation model, the weak disk only contributes a few soft photons to cool the corona, which in turn leads to harder X-ray spectrum.

\subsection{Jet in luminous AGNs}
 The jet is a very common feature in AGNs. However, the formation of jets is still a debated issue. Ballantyne (2007) investigated the accretion geometry of the radio-loud AGNs and suggested that there are three conditions affecting the jet launching: ``a rapidly spinning black hole, an accretion flow with a large ``H/R'' ratio,
 and a favorable magnetic field geometry''. In recent years, the observational work on the correlation index $\xi_{\rm RX}$ between the radio-luminosity $L_{\rm R}$ and hard X-ray luminosity $L_{\rm x}$ ($L_{\rm R}\propto L_{\rm x}^{\xi_{\rm RX}}$) has provided  more clues on the relation between the jet and the accretion flows (Merloni et al. 2003; Falcke et al. 2004; Wu et al. 2013). It is found that the formation of a jet is related to the hot plasma in the vicinity of the black hole, in the form of either ADAF at low accretion rates or a disk corona at high accretion rates (e.g., Yuan et al. 2008; Cao 2014; Huang et al. 2014; Sun et al. 2015; Gu et al. 2015; Zhang et al. 2015). These hot accretion flows are always geometrically thick with $H\sim R$. In our model, since the detailed configuration of the magnetic field is not clear, we neglect the jet/outflow escaping from the corona. If a rapidly spinning black hole exists in the center, we can suspect that this magnetic-energy-sustained corona might help in launching the jet in luminous AGNs. The jet formation may also help reduce the downward reprocessing in the cool disk, which will also affect the X-ray emission of the corona. This issue will be studied in the future work.

\section{CONCLUSION}
 We revisit the structure and the emergent spectrum of a disk corona heated by the reconnection of magnetic field
  generated in the disk by dynamo action and emerges into the corona.
We studied the effect of the magnetic field on the structure and spectrum of the model with various $\beta_{\rm 0}$ for $\alpha\,=\,0.3$. It is found that the thin disk changes from being radiation pressure dominated to being gas pressure dominated as the magnetic field increases, which smoothly join the two types of solutions together in Liu et al. (2002b, 2003). We find that the energy fraction gradually decreases from 0.95 to 0.3 when the accretion rate increases from 0.03 to 0.5 for $\beta_{\rm 0}\sim 100-200$. Correspondingly, the hard X-ray spectrum becomes softer at higher accretion rate, which is consistent with the observed results. However, the disk is dominated by the radiation pressure, and the corona is still energy inadequate for the hard X-ray emission with $\beta_{\rm 0}> 200$. For $\beta_{\rm 0}<50$, the disk is absolutely dominated by gas pressure and the energy fraction $f\sim 1.0$. The spectrum hardly changes with accretion rate, which is the same as the hard spectrum found in Liu et al.(2003).

\textbf{ACKNOWLEGEMENTS} We thank the referee for
very useful suggestions and comments. This work is supported by the National
Natural Science Foundation of China (Grant Nos. 11303046, 11303086, 1173026, and U1231203) and the Strategic Priority Research
Program ``The Emergence of Cosmological Structures'' of the Chinese Academy of Sciences (Grant No. XDB09000000) and   Gravitational Wave Pilot B (grant no. XDB23040100).




{}

\end{document}